\documentclass[a4paper]{jpconf}

\usepackage[english]{babel}
\usepackage{graphicx}
\usepackage[frozencache,cachedir=minted-cache]{minted} 
\usepackage[hidelinks]{hyperref}
\usepackage{orcidlink}
\usepackage{xspace}
\usepackage{booktabs}

\newcommand{\cpp}{C\nolinebreak\hspace{-.05em}\raisebox{.4ex}{\tiny\bf +}\nolinebreak\hspace{-.10em}\raisebox{.4ex}{\tiny\bf +}\xspace}
\xspaceaddexceptions{98 03 11 14 17 20 23 26}

\newmintinline{cpp}{}

\begin{document}

\title{Updates on the Low-Level Abstraction of\\Memory Access}

\date{2023-02-08}

\author{Bernhard Manfred Gruber\orcidlink{0000-0001-7848-1690}}

\address{EP-SFT, CERN, Geneva, Switzerland}
\address{Center for Advanced Systems Understanding (CASUS), Saxony, Germany}
\address{Helmholz-Zentrum Dresden-Rossendorf (HZDR), Dresden, Germany}
\address{Faculty of Computer Science, Technische Universität Dresden, Dresden, Germany}

\ead{bernhard.manfred.gruber@cern.ch}

\begin{abstract}
	Choosing the best memory layout for each hardware architecture is increasingly important
	as more and more programs become memory bound.
	For portable codes that run across heterogeneous hardware architectures,
	the choice of the memory layout for data structures is ideally decoupled from the rest of a program.
	The low-level abstraction of memory access (LLAMA)
	is a C++ library that provides a zero-runtime-overhead abstraction layer,
	underneath which memory mappings can be freely exchanged
	to customize data layouts, memory access and access instrumentation,
	focusing on multidimensional arrays of nested, structured data.
	After its scientific debut, several improvements and extensions have been added to LLAMA.
	This includes compile-time array extents for zero-memory-overhead views,
	support for computations during memory access,
	new mappings for bit-packing, switching types, byte-splitting,
	memory access instrumentation,
	and explicit SIMD support.
	This contribution provides an overview of recent developments in the LLAMA library.
\end{abstract}



\section{Introduction}

The performance gap between CPU and memory widens continuously --
many programs nowadays are memory-bound.
Compute and memory hardware is increasingly heterogeneous and
writing portable and performant programs becomes harder.
Memory-related optimizations typically depend on full control over data layout and memory access.
The Low-Level Abstraction of Memory Access (LLAMA) is being developed
as a portable, standard \cpp17/\cpp20 library to fill this gap~\cite{llama_github}.
At its core, LLAMA separates the algorithmic view of data from its mapping to memory,
allowing different data layouts to be chosen without touching the algorithm.

\begin{figure}
	\centering
	\includegraphics{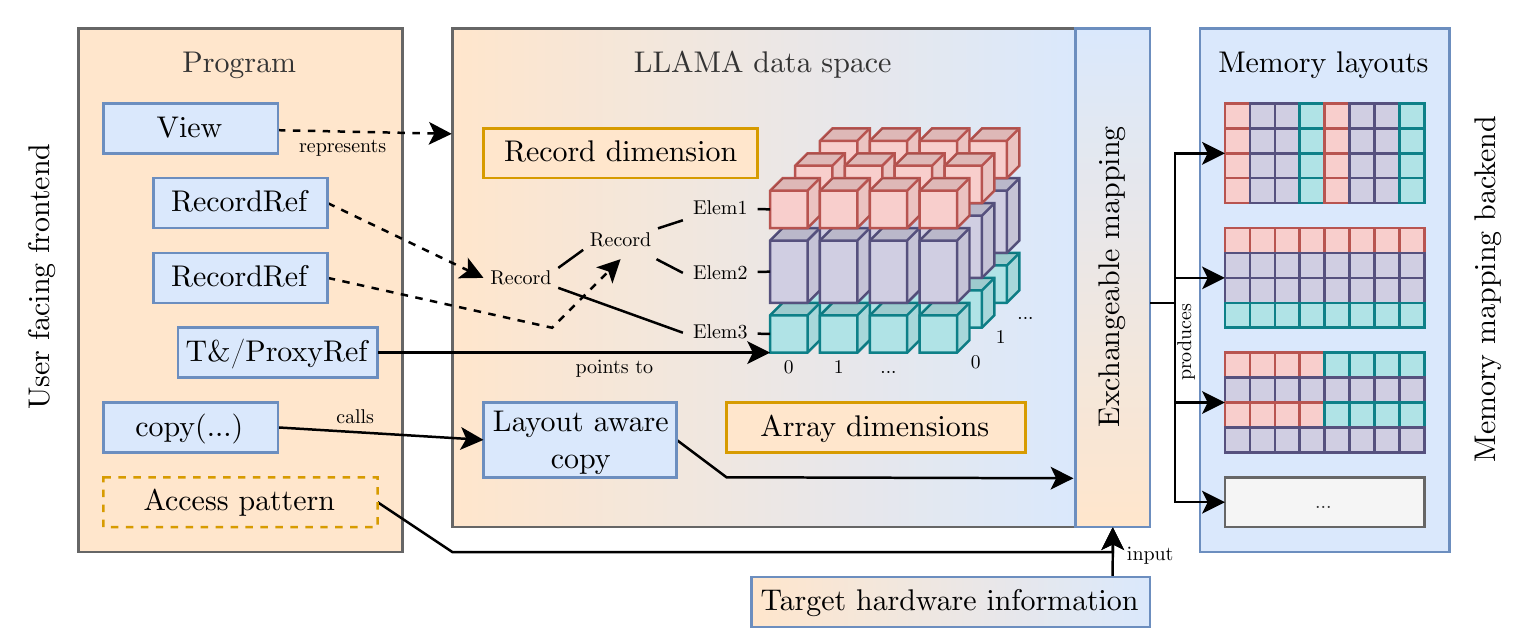}
	\caption{
		Conceptual overview of LLAMA.
		\label{fig:concept}
	}
\end{figure}

Conceptually, LLAMA uses a record dimension and several array dimensions
to span a data space of objects which should be mapped to memory.
A user's program interacts with this data space via a \cppinline|View|,
with individual records via \cppinline|RecordRef|
and with the final objects via l-value references or proxy references%
(a user-defined type that acts like a language built-in reference).
The data space is mapped to a memory layout using an exchangeable and user-definable mapping.
This mapping can be augmented with information on target hardware and access pattern.
LLAMA also supports layout-aware copy operations.
Despite these many capabilities, further extensions to LLAMA were necessary
and have been developed since its first publication~\cite{llama_paper}.
In this article, we would like to present these recently introduced features
and discuss their applications and use cases.

\section{Compile-time array extents}

Previously, the number of array dimensions in LLAMA were specified at compile-time,
but the extents of each dimension were strictly runtime values.
Also, all indexing and memory offset calculations used the \cppinline{std::size_t} data type,
which commonly has a 64-bit representation on many systems today.
This suffices for common CPUs and big LLAMA views,
where the array extents stored inside the view are negligible.
However, GPUs can incur higher costs for 64-bit integer arithmetic compared to 32-bit.
For example, dedicated 64-bit integer hardware is absent
from CUDA architectures like Hopper~\cite{H100_arch_blog} and the CUDA programming guide~\cite{cuda_prog_guide}.
And while the AMD MI200 architecture supports some 64-bit integer instructions,
it notable lacks arithmetic ones~\cite{MI200_isa_reference}.
Therefore, using smaller integral types is desirable.
Furthermore, small LLAMA views, e.g., placed in a GPU's shared memory,
cannot afford to additionally store the view's array extents.
Additionally, such extents are often derived from hardware quantities,
e.g., the shared memory size per streaming multiprocessor,
and known at compile time.

To account for these needs,
LLAMA now allows to specify the data type which should be used in all indexing computations.
Additionally, the array extents can be (partially) specified at compile time
and only runtime extents are stored.
If all extents are provided at compile time, array extents and mappings become stateless,
which is achieved by a careful implementation and use of empty-base-class-optimization~\cite{ecbo}.
Combined with the right blob allocator, the view becomes a trivial value type
and contains only the blobs for the mapped data.
The view is thus trivially constructible and storage-wise equivalent to the mapped data.
It can thus be memcpy-ed, reinterpret-casted from a buffer,
or placed in, e.g., CUDA shared memory.
Here are examples of the new array extents API:

\begin{minted}{cpp}
auto ae1 = llama::ArrayExtentsDynamic<int, 2>{size1, size2};
auto ae2 = llama::ArrayExtents<std::size_t, 3, llama::dyn, 4, 4>{size};
auto ae3 = llama::ArrayExtents<short, 32, 4, 4>{};
\end{minted}

The definition \cppinline{ae1} defines array extents with two dynamic sizes, using \cppinline{int} as index type.
Then, \cppinline{ae2} defines array extents with one static extent of 3,
a dynamic extent, and two more static extents of 4 each, using \cppinline{std::size_t} as index type.
Finally, \cppinline{ae3} defines fully static/compile-time extents und uses \cppinline{short} for all index arithmetic.
Allowing this mixing of compile and runtime extents was inspired by recent changes to the \cpp23 proposal \cppinline{std::mdspan}~\cite{mdspan}.


\section{New memory mappings}

Uses of LLAMA in real-world code bases and new environments led to the creation of new memory mappings,
lifted into the LLAMA library for general use.
In the following we would like to present the newly added mappings and their intended area of application:

\begin{description}
	\item[BitpackIntSoA and BitpackFloatSoA]
	Experimental data in high-energy physics is often taken using specialized hardware
	with a precision different than the \cpp standard fundamental types.
	Storing such values in the next bigger fundamental type wastes unnecessary bits of storage space
	but loads and stores use fast conventional hardware instructions.
	The BitpackIntSoA mapping allows to specify a desired bit count for integral types.
	The values will be packed/unpacked when stored to/loaded from memory.
	The bitpacked values are then further organized as Struct of Arrays (SoA),
	but we want to generalize this aspect in the future.
	The BitpackFloatSoA
	allows the user to individually specify the desired bit count for
	mantissa and exponent of floating-point values in the record dimension.
	Floating-point semantics are preserved as best as possible,
	including handling of NAN and INF values and mapping overflowing values during packing to INF.

	\item[Changetype]
	Because of the packing/unpacking overhead required by the bitpack mappings,
	a mere change of the storage data type is computationally more efficient,
	because the hardware may have appropriate conversion instructions.
	Such a conversion could, e.g., map a \cppinline|double| to a \cppinline|float|,
	or even to one of the \cpp23 extended floating point types~\cite{extended_fp_types}:
    \cppinline|float16|, \cppinline|float32|, \cppinline|float64|, \cppinline|float128| or \cppinline|bfloat16|.
	The adapted record dimension can then be mapped using a further mapping.
	This mapping was inspired from the accessor of Ginkgo~\cite{ginkgo_accessor}.

	\item[Bytesplit]
	Many compression algorithms are more efficient when compressing a stream of zeros.
	If the values in an integer array are small,
	the higher-order bytes may often be just zero.
	Splitting the values into their bytes and regrouping those by their order
	can effectively co-locate many zero-bytes
	and thus lead to higher compression ratios
    (cf. the \texttt{BYTE\_STREAM\_SPLIT} encoding in Apache Parquet).
	The Bytesplit mapping generalized this approach by splitting each type in the record dimension into a byte array of static size,
	and then forwarding the resulting record dimension to any further mapping.

	\item[Null]
	Sometimes we do not need to store all the fields of a LLAMA view.
	An example is a view acting as a cache to a different view,
	e.g., in GPU shared memory, for a particular algorithm that only works on a subset of the record dimension.
	A different use case is to remove the effect of accessing a field when, e.g., profiling.
	The Null mapping discards any values written to it and returns a default constructed value when reading from it.
	It is intended to be used together with the Split mapping,
	to select which part of the record dimension to not map to physical storage.

	\item[FieldAccessCount and Heatmap]
	These will be discussed in section~\ref{sec:instrumentation}.

\end{description}

While we have tested these new mappings in simple examples,
we would like to properly explore the impact of these new memory mappings in a future publication.

\section{Memory access instrumentation}
\label{sec:instrumentation}

LLAMA provides two instrumentation mappings, FieldAccessCount%
\footnote{The FieldAccessCount mapping was called Trace in previous versions of LLAMA.} 
and Heatmap.
The lightweight FieldAccessCount counts the accumulated number of accesses per record field.
The heavyweight Heatmap counts accesses to storage bytes at a configurable granularity such as bytes or cache lines.
Both mappings forward all mapping logic to, and can thus instrument, an arbitrary inner mapping.

Counting memory accesses is performed as side effect of data access
and costs one atomic increment to a dedicated memory location per regular access.
For CUDA, we measured, e.g., a ~3x slowdown in a particle transport simulation built with AdePT~\cite{adept_paper}.
%
While the size of the extra memory to store the counters is negligible for the FieldAccessCount mapping
(2 times the number of record fields),
the Heatmap at highest granularity requires an extra counter per byte of memory.
For a 64-bit (8 bytes) counter this results in an 8x memory overhead.

Software instrumentation, like discussed here, also comes with some limitations:
LLAMA cannot observe what the hardware and the compiler/optimizer do.
E.g., whether a memory read is served from RAM or cache,
or whether a second read to the same address is optimized out and served from a register.
Initial refactoring can help to increase the accuracy of instrumentation results,
e.g., by replacing repeated access to memory by a local variable.

We extensively demonstrate LLAMA's instrumentation capabilities in our integration into the AdePT project,
where we show tracing results and heatmaps of various memory access patterns
of a particle transport simulation~\cite{adept_llama_acat22_paper}.

\section{Explicit SIMD support}

Programs using Single Instruction Multiple Data (SIMD)
perform the same operation on $N$ operands at the same time,
typically supported by dedicated processor instruction sets.
Automatic vectorization of scalar code to SIMD instructions by modern compilers is brittle
and may fail for advanced codes,
requiring the use of explicit SIMD APIs and dedicated libraries~\cite{vc}.
LLAMA has been extended to support such SIMD libraries.

The primary interaction between SIMD types and LLAMA are memory-layout-aware $N$-element vector load and store operations.
Although many SIMD instructions also support memory operands in computational instructions,
those can usually be produced by the compiler by fusing a load/store and a compute instruction.
For convenience, LLAMA also provides simdized records,
a term adopted from the Vc library~\cite{vc}
meaning the creation of a SIMD version of a type.
Furthermore, load/store operations to handle scalar and simdized records uniformly have been added to LLAMA.
The API is independent of a SIMD library and integration is handled via type traits
(i.e., \cpp template specialization for user-defined types).


LLAMA can simdize scalar types or record dimensions (structured data) to a specified $N$
using the new \cppinline|SimdN| API,
as described in table~\ref{tbl:simdn}.
%
%
%
Algorithms should be written with a flexible $N$ to be portable.
However, $N$ needs to be fixed at compilation by the user
and depends on the target hardware, compilation flags and involved data types.
While this handles well-established SIMD instruction sets like AVX or NEON,
newer ones like ARM's Scalable Vector Extension have a runtime vector length.
We have yet to see how to deal with such instruction sets.
%
\begin{table}
	\centering
	\begin{tabular}{ l l l }
        \toprule
		\cppinline|SimdN<T, N, ...>| & \cppinline|N > 1| & \cppinline|N == 1| \\
        \midrule
		Record dim \cppinline|T|     & \cppinline|One<SimdizeN<T, N, ...>>| & \cppinline|One<T>| \\
		Scalar \cppinline|T|         & \cppinline|    SimdizeN<T, N, ...> | & \cppinline|    T | \\ \bottomrule
	\end{tabular}
	\caption{
		LLAMA's \cppinline|SimdN| creates SIMD versions of scalar types or records,
        with a desired SIMD width $N$.
        For $N = 1$, scalar types and records are created.
        Otherwise, \cppinline|SimdizeN| turns a scalar \cppinline|T| into a SIMD vector of \cppinline|T|,
        and a LLAMA record into a record with simdized field types.
		\label{tbl:simdn}
	}
\end{table}
%
%
%
Beside \cppinline|SimdN| to declare variables,
LLAMA offers the \cppinline|loadSimd| and \cppinline|storeSimd| functions
to transfer data between a SIMD construct or scalar and a reference to memory.
LLAMA will handle records and the underlying memory layout transparently for the user.
%
%
Figure~\ref{fig:nbody_update_simd} shows a simdized version of the update routine of the all-pairs n-body simulation
from the original LLAMA paper~\cite{llama_paper}.
With \mbox{$N > 1$} and the right compiler flags, SIMD code is produced.
For \mbox{$N = 1$}, a scalar version is generated without any trace of SIMD constructs.
The scalar version can run on CUDA.

\begin{figure}
	\centering
	\begin{minted}[gobble=1]{cpp}
	template <int N, typename ParticleView>
	void updateSimd(ParticleView& particleView) {
	  using Particle = ParticleView::RecordDim;
	  for(std::size_t i = 0; i < problemSize; i += N) {
	    llama::SimdN<Particle, N, std::fixed_size_simd> simdParticles;
	    llama::loadSimd(particleView(i), simdParticles);
	    for(std::size_t j = 0; j < problemSize; ++j)
	      pPInteraction(simdParticles, particleView(j));
	    llama::storeSimd(simdParticles(tag::Vel{}), particleView(i)(tag::Vel{}));
	  }
	}
	\end{minted}
	\caption{
		A SIMD version of the n-body update routine from the original LLAMA paper~\cite{llama_paper},
		using \cppinline{std::fixed_size_simd} as SIMD technology, as proposed for \cpp26~\cite{std_simd}.
		\label{fig:nbody_update_simd}
	}
\end{figure}

\begin{figure}
	\includegraphics[width=0.48\textwidth, trim=10 5 10 10]{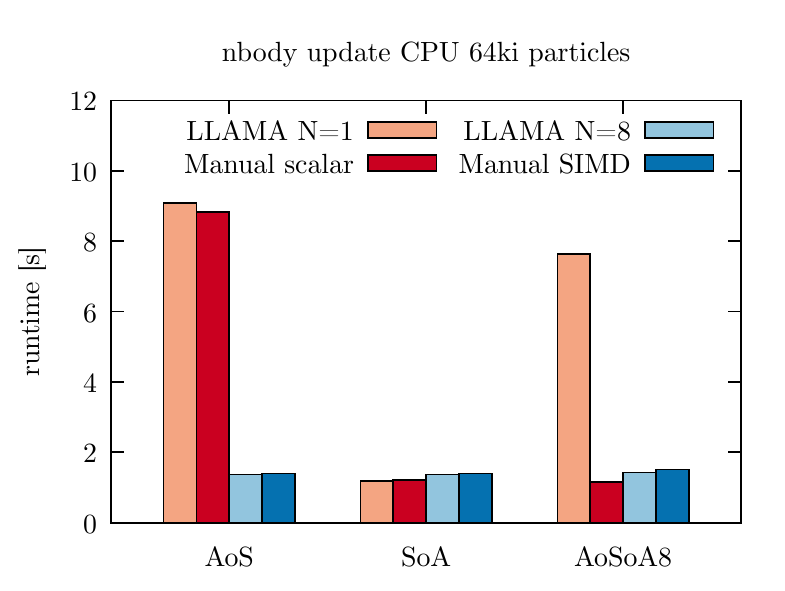}
    \hfil
	\includegraphics[width=0.48\textwidth, trim=10 5 10 10]{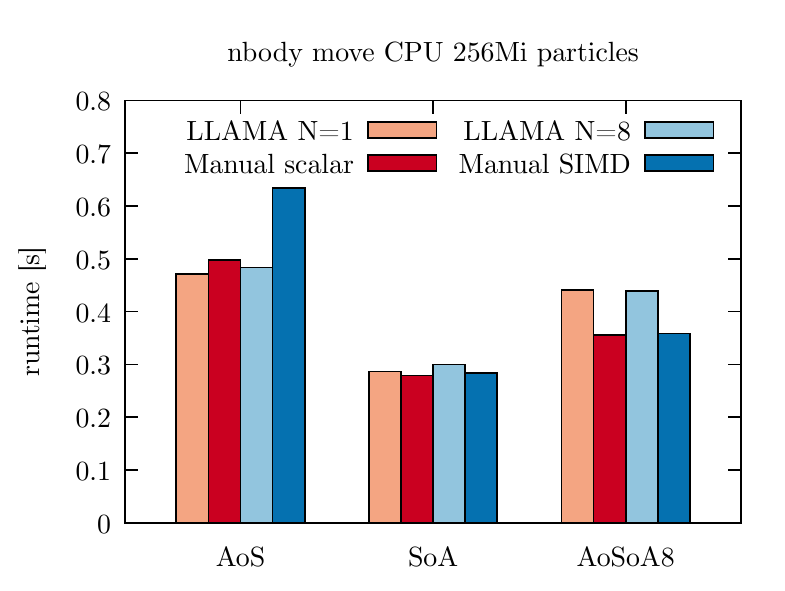}
	\caption{
		Benchmark of the CPU LLAMA n-body with a selection of popular mappings against various manually written versions.
		\label{fig:cpu_nbody_benchmark}
	}
\end{figure}

Figure~\ref{fig:cpu_nbody_benchmark} compares a LLAMA n-body simulation against manually written scalar and SIMD versions
on an AMD Ryzen 9 5950X CPU with AVX2.
Results are single-threaded to emphasize the efficiency of the generated instructions.
The scalar runs with the Arrays of Structs (AoS) layout are not auto-vectorized by the compiler.
LLAMA matches the manually written code here.
The manual SIMD implementation of the move step for the AoS layout uses gather instructions,
whereas LLAMA uses multiple scalar loads,
for which the compiler seemed to generate better code for the target CPU.
Replacing the gather instructions in the manual SIMD move by multiple scalar loads,
gets the runtime on par with the LLAMA SIMD runtime.
For SoA, the multi-blob (MB) version is used, which stores each field in a separate allocation,
and the results are nearly on-par between LLAMA and manually written versions.
Both scalar codes have been auto-vectorized by the compiler.
The Arrays of Structs of Arrays (AoSoA) layout in LLAMA has overhead in this example,
which is especially visible in the n-body move phase
and explained in more detail in the LLAMA paper~\cite{llama_paper}.
It is caused by the LLAMA version using a single for-loop to traverse the array index space once,
while the manual AoSoA version can use two nested for-loops (traversing AoSoA blocks and lanes),
matching the memory layout structure and allowing easier auto-vectorization.
Further investigation is pending to provide such mapping-aware loop structures inside LLAMA as well.
%
Except for the AoSoA, LLAMA generally fulfills the zero-overhead principle in scalar and SIMD code.

\section{Summary and outlook}

We have presented recent updates and new features of LLAMA since its previous publication.
Compile-time specification of array extents support common use cases for shared memory caches in GPGPU programming.
New memory mappings open advanced possibilities for separating arithmetic
from in-memory precision with various tradeoffs,
supporting data arrangement for improved compression,
and enabling instrumentation and tracing of memory access patterns.
Finally, explicit SIMD support for LLAMA closes the gap between SIMD computing,
structured data and arbitrary memory layouts.

In the future, we will focus on testing LLAMA with more real-world applications and on more hardware platforms.
This will include the development of a systematic workflow
to improve memory-related performance aspects of such applications.
Supporting additional access patterns beyond LLAMA's random access could further solve the slow AoSoA
and pave the way for LLAMA mappings with block compression algorithms.

\ack

This work has been sponsored by the Wolfgang Gentner Programme of the German Federal Ministry of Education and Research (grant no. 13E18CHA).
The author would like to thank Guilherme Amadio, Verena Gruber and Stephan Hageböck for proof-reading and commentary.

\section*{References}

\bibliographystyle{iopart-num-with-article-titles}
\bibliography{bibliography}

\end{document}